# Parallelized aperture synthesis using multi-aperture Fourier ptychographic microscopy

**PAVAN CHANDRA KONDA, JONATHAN M. TAYLOR, ANDREW R. HARVEY**[*]

*Imaging Concepts Group, School of Physics and Astronomy, University of Glasgow, Scotland, G12 8QQ, UK*
*Corresponding author: Andy.Harvey@glasgow.ac.uk*

We report a novel microscopy platform, termed Multi-Aperture Fourier ptychographic microscopy (MA-FPM), capable of realizing gigapixel complex field images with large data acquisition bandwidths – in gigapixels per second. MA-FPM is a synthetic aperture technique: an array of objectives together with tilt-shift illumination are used to synthesize high-resolution, wide field-of-view images. Here, the phase is recovered using Fourier ptychography (FP) algorithms, unlike conventional optical synthetic aperture techniques where holographic measurements are used. The parallel data-acquisition capability due to multiple objectives provides unprecedented bandwidth enabling imaging of high-speed *in vitro* processes over extended depth-of-field. Here, we present a proof-of-concept experiment demonstrating high-quality imaging performance despite using nine-fold fewer illumination angles compared to an equivalent FP setup. Calibration procedures and reconstruction algorithm were developed to address the challenges of multiple imaging systems. Our technique provides a new imaging tool to study cell cultures, offering new insights into these processes.

## 1. INTRODUCTION

Perhaps the most important advantage of the emerging technique of Fourier ptychography (FP) [1] is the ability to record gigapixel images with sub-micron resolution using low numerical-aperture (NA) optics. In particular, FPM provides wide-field, high-resolution imaging without mechanical scanning. Instead, an extended spatial-frequency spectrum of the image is assembled in time-sequence [2] and computationally integrated into a single high-bandwidth complex spatial-frequency spectrum of the image from which a high-resolution complex image is calculated [3,4]; i.e., the optical phase of the sample is also reconstructed. The enhanced resolution is achieved through synthesis of a high-NA illumination source; the high-frequency cutoff may then be equivalent to that of a higher-NA objective, but the recording and reconstruction process provides the greater depth of field and field of view associated with the low-NA objective used to record the band-pass images. These numerous advantages are offset, however, by the reduced recording speed associated with time-sequential synthesis of the high-NA illumination; a factor that is exacerbated by the requirement for some redundancy in the sampling of overlapping spatial-frequency bands [5–7].

In conventional microscopy the objective acts as a low-pass filter and the higher diffraction orders, familiar from the Abbe theory of image formation are lost, reducing both the resolution and bandwidth of acquired images [8]. In FP, the objective does not act as a low-pass filter, but instead as a bandpass filter – but nevertheless all spatial frequencies not transmitted by the objective are lost. Since several bandpass images with differing spatial-frequency content are recorded in FP, a large space-bandwidth product (SBP) [7,9] is achieved with the *quid pro quo* of increased recording time. We report here the first demonstration of multi-aperture Fourier ptychographic microscopy (MA-FPM), in which multiple objectives, each forming an image on an independent detector array, capture this diffracted light and hence increase the space-bandwidth-time product (SBTP) [7,10] for image acquisition by a factor equal to the number of objectives used. In our demonstration we employ nine cameras in parallel to yield a nine-fold reduction in the number of time-exposures required to form a high-resolution image.

The high-NA illumination is implemented using an array of programmable light-emitting diodes (LEDs) [11–13] where each individual LED illuminates the object at a distinct angle, resulting in a shift of spatial frequencies sampled by the objective according to Ewald-sphere theory [14,15]. Band-limited images are stitched together in frequency space using alternate-projection type algorithms [16–18] to reconstruct the high-resolution optical field at the object. Since the complex field of the object is reconstructed, this technique can also provide dark-field and phase-contrast images of the object [19,20]. The image-reconstruction algorithms also facilitate estimation and correction of unknown aberrations in the optical system [4,21].

Fundamental to MA-FPM is the recognition that multiple apertures can record multiple bands of high-spatial-frequency diffracted light in parallel enabling coherent reconstruction of the high-bandwidth complex amplitudes of the diffracted fields [22]. MA-FPM involves mutually incoherent recording of the spatial-frequency bands transmitted by each aperture combined with algorithms that achieve the equivalent function of arrays of apertures that coherently sample extended fields.

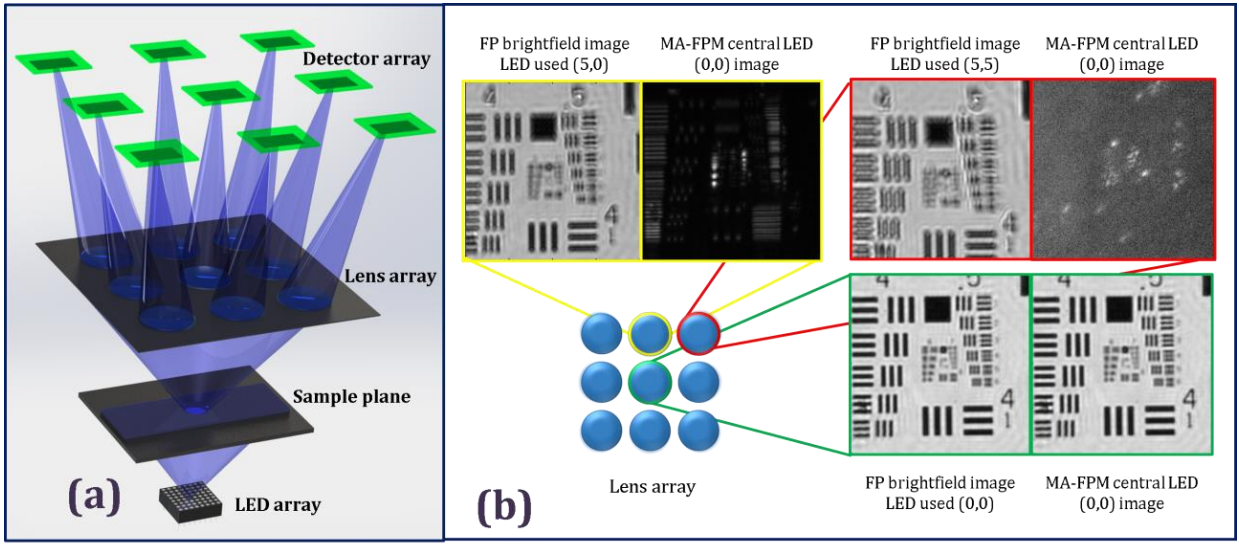

**Fig. 1. (a)** Planar MA-FPM setup. **(b)** Images obtained by the lenses in the array: Three lenses at different positions with various magnitudes of aberrations are chosen to demonstrate the differences in images and the spatial frequency content imaged. Images in the green box are from the central lens with zero off-axis aberrations, images in the yellow box are from a vertical lens which has off-axis aberrations in only one direction, images in red box are from a diagonal lens which has off-axis aberrations in both vertical and horizontal directions. In each set, images on the left are bright-field images where each lens requires different LED illumination to capture bright-field frequencies. Here it can be clearly seen that these bright-field images are mutually translated due to alignment errors and contain dissimilar aberrations associated with their off-axis positions. Images on the right are each obtained using the same central LED. Here each lens will record different set of spatial frequencies as observed.

Consequently MA-FPM synthesizes the enhanced resolution and bandwidth of a single larger synthetic aperture, a concept widely used in radio-frequency imaging employed in astronomy and remote sensing for high-resolution imaging [23]. A key distinction however is that for radio frequency aperture synthesis, the electromagnetic field is directly sampled and correlated in the pupil plane with single-mode receivers, whereas in MA-FPM the field is sampled in the pupil apertures focusing onto detector arrays. In both cases redundancy in sampling is required to yield efficient complex calibration and image reconstruction.

Optical aperture synthesis has previously been reported in microscopy [24–27] but requires the complexity and sensitivity of optical interferometry for phase recovery and is limited to a single imaging aperture. Conversely, our use of FP-based iterative computational calculation of diffracted complex fields, sampled by multiple apertures, enables a flexible combination of time-sequential and snapshot imaging to yield both increased spatial resolution and reduced recording time. The experimental complexity required for conventional interferometric aperture synthesis can be considered to be exchanged for the increased computational complexity required to compute the final image. This modern approach to imaging exploits the exponential increase in the power of low-cost computing.

In MA-FPM, the increased space-bandwidth product recorded in a single snapshot enables a reduction in the number of frames and total integration time required to construct a high-resolution image. That is, by using an $n \times n$ array of objective lenses the number of LED-illuminations can be reduced by a factor of $n^2$, providing also a factor $n^2$ reduction in acquisition time. A further factor-of-eight reduction in acquisition is possible (for the optimal 65% overlap of spatial frequencies) by employing coded, multiplexed LED illumination [10,28–30] with sparse multiple objectives. It is noteworthy that if dense sampling by multiple objectives is implemented, MA-FPM provides a route to snapshot operation using an optimized array of only nine LEDs.

The lens mounting structures of practical multi-objective arrays result in gaps between lens apertures with associated gaps in frequency coverage for each acquisition. Hence, it is necessary for the design of the objective array and the LED array combination to permit recording of all spatial-frequency bands and retain the redundancy required for FP reconstruction [5,12,31]. The MA-FPM configuration depicted in Fig. 1(a) achieves this by matching the LED array size with the gaps between the lenses. To implement MA-FPM in microscopy, the off-axis geometry associated with large object-space NA means that it is necessary to employ custom-mounted discrete objective lenses rather than off-the-shelf integrated cameras as has been reported for macroscopic multi-camera FP [32]. Due to variations in geometry and manufacturing tolerance, each lens-detector pair in MA-FPM exhibits distinct aberrations and magnification, which we calibrate independently for optimization of image quality. In these circumstances the Fraunhofer approximation for propagation of light fields, which is conventionally employed in FP, cannot be efficiently employed [22,33]. We have therefore developed and applied a new reconstruction procedure based on Fresnel propagation [21,34,35] that enables correction for camera-to-camera variations in magnification and aberrations yielding a higher-quality reconstruction.

Prior to Fourier ptychographic stitching of the spatial-frequency spectra of the multiple images into a single high-resolution spectrum, co-registration of all recorded images is necessary. These images are dissimilar due to geometrical distortions and imaging aberrations and, of course, due to the differing image characteristics of the different spatial- recorded frequency bands (for example dark-field and bright field images are very dissimilar due to the corresponding high-pass filter and low-pass filter functions), and so conventional image registration algorithms [36,37] are ineffective. Our calibration procedure, described below, solves these problems to yield a high-quality image with an overall space-bandwidth-time product (SBTP, i.e., the product of the frame rate and the space-bandwidth product for a single acquisition) that is much greater than that of a single camera.

Multi-camera imaging systems are also attracting increasing interest in macroscopic imaging; for example to super-resolve aliased images and to enable increased functionalities such as foveal imaging and multi-spectral imaging [38–43]. The essential motivation is to introduce and exploit diversity to increase the data bandwidth through parallelized image recording: in spatial sampling, in magnification and in spectral bandpass in the above examples. These earlier multi-camera techniques involve incoherently combining multiple images, and hence cannot be used to recover phase and are limited to macroscopic imaging. However, our introduction of multi-camera imaging to Fourier-ptychographic microscopy is a coherent, aperture-synthetic combination of the complex spatial-frequency spectra. Hence, this is the first time multiple cameras are used in microscopy to provide a route to a disruptive increase in the SBTP (here we demonstrate an order-of-magnitude improvement).

In the next section we present two possible MA-FPM configurations, discuss their feasibility and then present the calibration employed together with the image-reconstruction procedure developed for MA-FPM. We present experimental verification of MA-FPM involving translation of a single detector and lens system as required for a planar MA-FPM design. Geometrical relationships in the experimental configurations and detailed results from the calibration procedure are presented in the supplementary material S1 along with details of the Fresnel propagations used in the reconstruction algorithm.

## 2. MA-FPM setup and methods

As highlighted above, FP may employ a simple, low-NA objective lens instead of a sophisticated, high-cost high-NA objective [44–46] to provide a route to low-cost high-throughput gigapixel microscopy. The MA-FPM system consists of an array of simple lenses, each sampling a distinct angular range of the light diffracted by the object, to form band-limited images at an array of detectors as depicted in Fig. 1a. The effective NA for imaging is $NA_{eff} = NA_{obj} + NA_{ill}$, where $NA_{obj}$ is the synthesised object-space NA and $NA_{ill}$ is the synthesised illumination NA. Resolution therefore increases with both the number of objectives and the number of LEDs. Equivalently, an increased frame rate for a given SBP can be achieved by increasing the number of objectives while reducing the number of LED illuminations; for example, the effective NA that is achievable using a conventional single-objective FP system with 15x15 LEDs can be achieved by either 9 cameras and 5x5 LEDs or 25 cameras and 3x3 LEDs. For these two equivalent examples, multi-aperture FP offers a 9-fold and 25-fold increase in SBTP respectively compared to conventional FP. It should be noted that the spatial separation of the objectives in the first case would be larger and combined with a larger number of LED illuminations to provide the required spatial-frequency range. The geometry of the MA-FPM is chosen such that there is a minimum of 50% overlap between the spatial-frequency bands sampled by adjacent LEDs [5,31].

In the case of implementing LED multiplexing, an LED array with a minimum of 5x5 LEDs is desired to provide the degree of irregularity required in the multiplexed patterns [28]. FP is generally conducted in air without the benefit of oil immersion; the theoretical maximum NA is therefore limited to $NA_{eff} < 1 + NA_{obj}$. For MA-FPM, the $NA_{obj}$ can be increased to unity synthetically using multiple apertures enabling a larger $NA_{eff}$. MA-FPM therefore provides the unique advantage over conventional FP of being able to attain high $NA_{eff}$ (e.g. approaching two) using low-NA objectives, while retaining the low-NA benefits of large FoV and DoF.

Fig. 1b shows images recorded by three of the nine MA-FPM cameras: each camera may record both bright-field and dark-field images depending on the LED used for illumination. The bright-field images (on the left of each box), recorded with the appropriate co-axial LEDs, are mutually translated due to relative misalignment and are subject to dissimilar optical aberrations. For example, images recorded using the top-right objective are subject to the highest levels of off-axis aberrations (principally coma and astigmatism), while images recorded using the central and on-axis objectives are subject to much lower aberrations. Conventional image-registration algorithms are therefore ineffective. For illumination by only the central LED of the 5x5 LED array, only the central camera of the 3x3 lens array, located at position (2, 2), records a bright-field image, while the other eight cameras record dissimilar dark-field images. Three of these images are shown on the right of each box: the central objective records low spatial frequencies (the bright-field highlighted green) whereas the objective at (2, 1), and diagonal lens at (3, 1), record high-band, dark-field spatial frequencies.

### A. Planar MA-FPM

The planar MA-FPM concept, as shown in Fig. 1a, consists of a planar array of objectives forming an array of images of the object on a parallel planar array of detectors. The array of detectors can be translated axially in unison for focusing. Due to the aberrations arising from off-axis imaging (as are apparent in Fig. 1b and are discussed above) the quality of the final high-resolution image reconstruction is reduced and so this configuration is less suitable for very high-NA systems. This planar MA-FPM configuration provides, however, a proof-of-concept demonstration. We describe next a preferred geometry employing the Scheimpflug configuration for low-aberration, high-NA, off-axis imaging. The geometrical relations for planar MA-FPM and system parameters calculations are presented in the Supplementary material S1.

### B. Scheimpflug MA-FPM system

To reduce the aberrations associated with off-axis imaging we propose the Scheimpflug configuration [47,48], such as was initially developed for correcting perspective distortion when imaging tilted scenes, as depicted in Fig. 2. According to the Scheimpflug principle, the detector plane should pass through the intersection point of the object plane and the lens plane as shown in Fig. 2. In this configuration, perspective distortion is compensated by the varying magnification across the scene resulting in a low-distortion image. This configuration is used for off-axis lens systems in MA-FPM to correct for perspective distortion and to reduce the field-varying defocus and off-axis aberrations.

This Scheimpflug configuration does not suffer from the gross off-axis aberrations arising from off-axis imaging; hence, it offers the advantage over the planar configuration of being scalable to high $NA_{obj}$ and $NA_{eff}$. There are however some practical challenges: for example, focusing requires optimisation with three degrees of freedom for orientation of the lens and the detector. This can be solved by custom designing and 3D printed the setup. A prototype Scheimpflug MA-FPM is currently under construction and will be reported in a subsequent publication.

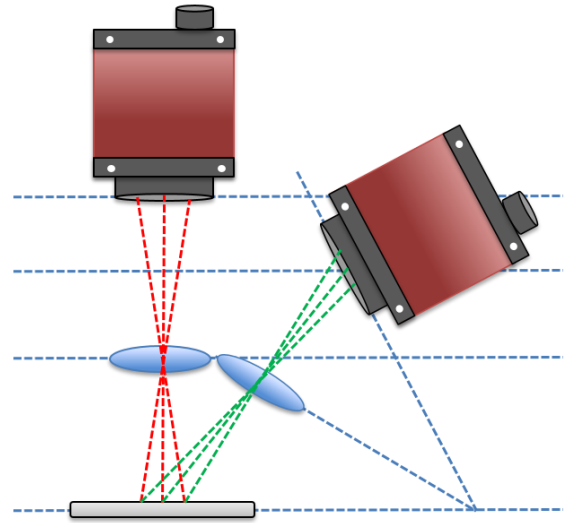

**Fig. 2.** Scheimpflug MA-FPM geometry showing the positions of imaging lenses and detectors.

### C. Calibration procedure

High-quality, artefact-free image recovery requires calibration of the system characteristics, in particular the pupil function (aperture shape and aberrations) and LED positions [3]. Additional factors specific to MA-FPM are camera-to-camera image registration and magnification. Magnification can be easily calibrated by imaging an object of known dimensions, but the image registration represents a greater challenge.

The variation in quality of the bright-field images apparent in Fig. 1b precludes the use of conventional image-registration algorithms. In our approach, we record calibration data for each individual lens system employing LED illumination such that each camera records the same spatial frequency band and hence can be mutually registered. Conventional single-objective FP images, with $NA_{FP} = NA_{obj} + NA_{ill}$, are constructed from these datasets for each lens using the FP-reconstruction procedure described in [28]. These images differ only in the translation arising from the uncertainty in detector positioning. An appropriate translation matrix is calculated by minimizing the error of the difference of these images to mutually align the high-NA images. These images can be aligned with an accuracy better than that of the recorded images due to the higher pixel count of the constructed images.

The pupil function for each lens can be recovered during the FPM reconstruction due to the redundancy present in the acquired data [4] and this enables mitigation of imaging aberrations. The redundancy in the MA-FPM data captured is less than FPM data since fewer images are recorded for each lens, but we compensate for this by increasing the redundancy in the calibration data by use of a larger number of LED positions than are used in image reconstruction. To further enhance the speed and robustness of convergence of the pupil-estimation algorithm we provide an initial estimate of the aberrations using a ray-traced model implemented using *Zemax®*. This calibration of the pupil function yields high-quality reconstruction as can be seen from the results presented later in Fig. 4 below. Errors in the LED positions can also be corrected if required by the techniques used in [49,50] on the calibration datasets. More details and results on the calibration datasets can be found in the supplementary material S1.

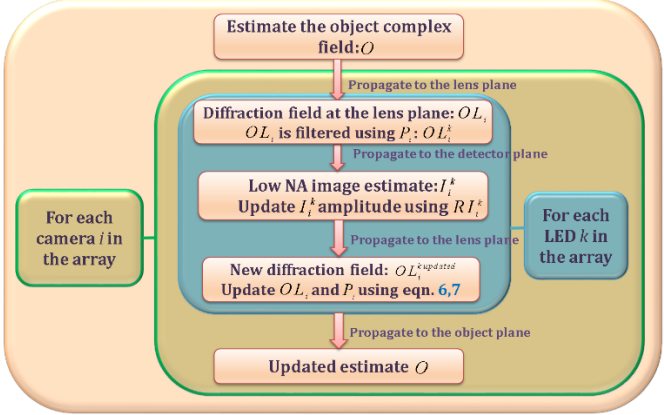

**Fig. 3.** Flowchart describing MA-FPM reconstruction algorithm. $O$ is the high-resolution complex field of the object that is to be recovered, $OL_i$ is the complex field due to the object in lens plane $i$, $OL_i^k$ is the band-pass filtered complex field of the spatial-frequency band corresponding to the LED illumination angle $k$ using lens $i$, $P_i$ is the pupil function of the lens $i$, $I_i^k$ is the estimated complex field on the detector plane for illumination by the LED at angle $k$ using lens $i$, $RI_i^k$ is the experimentally recorded intensity image for LED illumination $k$ using lens $i$ and $OL_i^{k\,updated}$ is the updated complex field of $OL_i^k$.

### D. Reconstruction procedures

In a single camera FPM system, the magnification and associated aberrations are identical for all recorded low-NA images. For a MA-FPM system however, aberrations and magnifications vary from camera to camera as described above, and the need for accurate correction will become more severe with increasing $NA_{obj}$ and $NA_{MA-FPM}$, due to greater sensitivity to alignment errors. Stitching of images with dissimilar magnification using conventional FP algorithms requires image interpolation, which can in principle introduce additional noise into the images already suffering from aberrations and low SNR. Hence, we have developed an algorithm based on Fresnel propagation which enables stitching of images with dissimilar magnifications.

Data acquisition proceeds as follows. Each LED in the array is illuminated in time sequence and the cameras record a distinct spatial-frequency band specific to the LED position and the camera position. In this demonstration the single detector is scanned through a matrix of positions to simulate the array of objectives. The images from the array of camera positions are registered as described in the previous section and pupil functions for each of these lenses are obtained from the registration step. These registered images and the estimated pupil functions are used to calculate the synthetic image with a resolution associated with $NA_{MA-FPM}$ using the reconstruction procedure described below.

MA-FPM reconstruction follows a process similar to FPM reconstruction procedures in [4,28] with a few generalizations to accommodate dissimilar pupil functions of the multiple camera positions and magnifications as summarized in the flowchart in Fig. 3. The reconstruction procedure commences by estimating the high-resolution complex optical field at the object, $O$, by interpolation of a low-NA image ($NA=NA_{obj}$). This object field has a factor eight more pixels than the corresponding image in each dimension in the detector plane (determined by the value of $NA_{MA-FPM}$ [1]), and the size of each pixel in the object field is a factor $8M$ smaller than the pixel size on the detector plane, where $M$ is the magnification of the system. The factor of 8 arises from the increase in the frequency bandwidth and hence pixel count achieved by FP. The complex fields $OL_i$ at each lens plane is then calculated in sequence using the Fresnel approximation and taking into account the specific geometry for each lens as described in the supplementary material S1.

The complex field $OL_i$ at each lens $i$ is filtered using the pupil function, $P_i$ of the lens, shifted according to the illumination by LED $k$ to obtain the filtered complex field, $OL_i^k$ at the output of the lens plane. Propagation of $OL_i^k$ to the detector plane yields $I_i^k$; an estimate of the image for illumination angle $k$. The amplitude of this estimated image is known from the recorded intensity image $RI_i^k$. Hence the amplitude of $I_i^k$ is updated using this information and the phase is retained. This is now propagated back to the lens plane to obtain an updated estimate of the complex field $OL_i^{k\,updated}$ which contains only those spatial frequencies corresponding to the $k^{th}$ LED illumination angle. This spatial-frequency information in the $OL_i^{k\,updated}$ is used to update the corresponding spatial-frequency band $OL_i$ using the second-order Gauss-Newton method described in [3]:

$$OL_i^{updated} = OL_i + \frac{|P_i| \cdot [OL_i^{k\,updated} - OL_i^k]}{|P_i|_{max} \cdot (|P_i|^2 + \alpha)}, \qquad (1)$$

and the pupil function of the lens $P_i$ is updated using

$$P_i^{updated} = P_i + \frac{|OL_i| \cdot [OL_i^{k\,updated} - OL_i^k]}{|OL_i|_{max} \cdot (|OL_i|^2 + \beta)}, \qquad (2)$$

where $\alpha$ and $\beta$ are regularization constants.

As shown in Fig. 3, this process is repeated for all LEDs in the array for each lens $i$. The new complex field $OL_i$ is then propagated back to the object plane to obtain an updated estimate of the high-resolution complex field $O$, which is used to calculate the diffracted field at the next lens and the whole process is repeated for the next lens. This process is repeated for all the lenses which completes one iteration of the reconstruction procedure. The whole process is iterated several times until the RMS error between two successive iterations falls below a required criterion; typically, convergence is achieved for between fifteen and forty iterations, depending on scene characteristics.

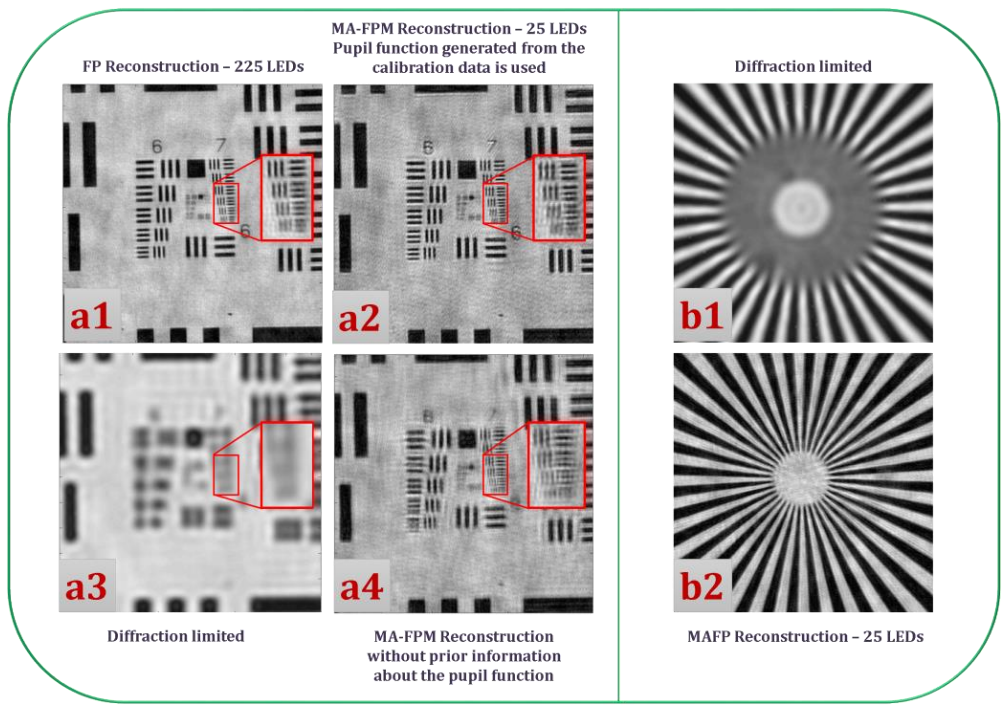

**Fig. 4.** MA-FPM validation results: (a1-a4) USAF resolution chart imaged to assess the resolution improvement quantitatively. a1– 225 LEDs FP reconstruction, a2 – 25 LEDs 9 cameras MA-FPM reconstruction using calibrated data and pupil phase estimated from calibration, a3 – diffraction limited image, a4 – MA-FPM reconstruction with calibrated data and without any pupil phase estimate (Most of the group 7 and some group 6 elements are not reconstructed properly). (b1-b2) A spoke target is imaged to show uniform improvement in the resolution in all directions, demonstrating that spatial frequencies in all directions are reconstructed.

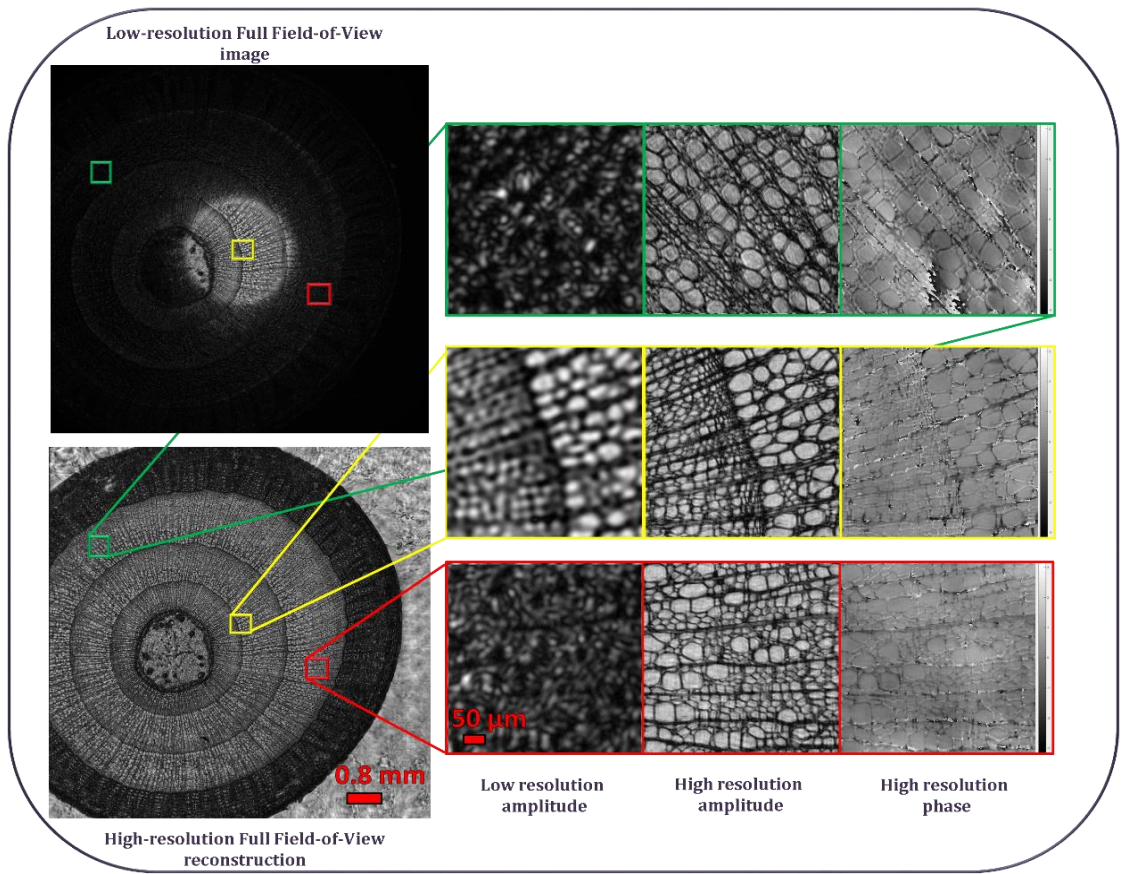

**Fig. 5.** Full FoV high-resolution reconstruction: *Brunel Woody dicotyled-stem cross section taken at 3 years* microscope slide imaged using MA-FPM. 3 different parts of the FoV are chosen to demonstrate the reconstruction quality across the FoV. Yellow box is the central section, red box is 30% of the FoV from center and green box is at 75% of the FoV from the center. In each box image on the left is the diffraction limited image, the central image shows the reconstructed high-resolution amplitude, and the right image is the reconstructed high-resolution phase.

## 3. Experiment and Results

We demonstrate an experimental implementation of a planar MA-FPM system consisting of a 3x3 lens array with illumination provided by a 5x5 LED array. This provides equivalent resolution and space-bandwidth product to a conventional FP system with a 15x15 LED array, but with a nine-fold increase in data acquisition bandwidth. The objective lenses are simple achromats with focal lengths of 36mm and aperture diameters of 3mm (*Edmund Optics #47-655*) giving an NA of 0.025 and a magnification of 1.7. An LED array (*Adafruit* P4) with LEDs arranged on a 4-mm period matrix was placed 257mm from the object plane to yield an overlap of 61% in the spatial-frequency bands recorded using adjacent LEDs. This configuration provides either a synthetic $NA_{FP}$ = 0.119 when using a 15x15 LED array or an equivalent $NA_{MA-FPM}$ = 0.119 using a 5x5 LED array, representing a factor of four increase in the NA of the objectives. The center-to-center separation between the lenses in the array is 4.25mm such that when the fifth LED from the center illuminates the object a bright-field image can be seen on the off-axis lens system. The detector used was an *Andor Zyla* 5.5 (16-bit sCMOS with 6.5µm pixel size.) For convenience, in this initial proof-of-concept experiment, MA-FPM images were recorded sequentially using a single lens and camera system translated through a rectangular array of positions simulating an MA-FPM system. Exposure times were 100 milliseconds for central camera images and 1000 milliseconds for rest of the cameras, and all images were normalized with respect to exposure time. The reconstructed images presented below were recorded using the red LEDs of the LED array (center wavelength 623nm and full-width at half maximum of 17nm). For reconstruction, the 2048x2048-pixel image on the detector is divided into sub-image segments of 256x256 pixels and processed independently to satisfy the partial-coherence requirements [22].

In Fig. 4 we present results for our MA-FPM system and an equivalent conventional FP system, which show similar image quality, demonstrating the validity of the techniques used to construct the image. MA-FPM images reconstructed with and without calibration are presented, which are indicative of the importance of calibration in achieving high-quality images. Fig. 4. a1-a4 shows images of a USAF resolution chart that demonstrates increase in resolution with increasing NA. The low-resolution bright-field image recorded by a single objective in the center of the array is shown in a3. The finest resolvable detail is group 5 element 4, corresponding to diffraction-limited imaging with $NA_{obj}$=0.025. In image a1, recovered using conventional FP employing 225 LEDs, group 7 elements 4 and 5 are clearly resolved indicating the expected $NA_{FP}$=0.119. Image a2 is recovered using our MA-FPM technique and also resolves group 7 elements 4 and 5 and with an equivalent $NA_{MA-FPM}$ = 0.119 to that achieved with conventional FP. These results confirm that MA-FPM reconstruction using the calibration procedure described above compares well to an equivalent FPM system, while offering a nine-fold increase in data-acquisition bandwidth. Fig. 4 b1-b2 presents a spoke target imaged using our MA-FPM system that demonstrates the omnidirectional resolution improvement.

Though care is taken to align the system precisely to specifications of the design there is inevitably some misalignment due to construction tolerances. Hence the actual phase aberrations of a practical system deviate from those estimated from the ray-trace model (implemented with *Zemax*®) and this is apparent from the calibration results shown in supplementary material S1. Hence using the aberration estimated using the calibration procedure gives a better reconstruction quality and faster convergence. In Fig. 4 a4 the reconstruction quality is degraded since the calibrated pupil phase estimate is not used in the reconstruction. FP reconstruction algorithms converge very slowly when the pupil function has large aberrations, thus an initial estimate of the aberrations improve the convergence speed and also yields better image quality. Furthermore, for conventional FP, the pupil function is identical for all LED illuminations (225 in the case considered here) yielding a higher convergence rate than for MA-FPM (which uses fewer images per pupil: 25 in our example). An accurate initial estimate of pupil aberrations is therefore likely to be more important for MA-FPM than for FP. Our calibration allows us to achieve this by finding the best estimate of the system aberrations.

A full-FoV reconstruction of an MA-FPM image is shown in Fig. 5: an image of a *Brunel* microscope slide of *woody dicotyled-stem cross section taken at 3 years*. It can be clearly seen that the recovered image exhibits very high resolution for the whole microscope slide. Three localized areas across the FoV are magnified to show the smallest features (cells of various sizes) that are recovered. These three image segments are taken from the central section of the FoV (highlighted in yellow), 30% of the FoV away from the center (shown in red) and 75% of the FoV from the center (shown in green) to demonstrate the high-quality reconstruction across a wide FoV. The reconstruction quality in the edges (over 75% FoV) is degraded slightly due to the higher levels of aberrations present in the edges of the FoV. These aberrations will be reduced experimentally using the Scheimpflug MA-FPM setup depicted in Fig. 2. We also present the recovered phase of these segments on the right. The phase information recovered can be very useful to determine the structure of transparent samples without staining [10]. In the current sample the phase information can be useful to find the thickness and axial positioning of the cells.

## 4. Conclusion

We have demonstrated a new aperture-synthetic MA-FPM microscopy technique with a SBTP of many gigapixels per second and can be used to implement several imaging modalities. Our implementation uses nine cameras to increase data acquisition speed by a factor of nine compared to a conventional FP system. We report new image-reconstruction algorithms and high-quality image recovery in this new microscopy architecture, and calibration procedures to register images from multiple cameras with sub-pixel accuracies. The use of aberrations estimated from the calibration procedure enable faster convergence and improved reconstruction. The resultant images generated using MA-FPM are of comparable quality to a conventional FPM system but are obtained with a factor of 9 fewer sequential LED illuminations. We have proposed a Scheimpflug MA-FPM system to reduce the magnitude of off-axis aberrations and promise higher image quality. A Scheimpflug MA-FPM system constructed from 3D-printed components and low-cost sensors is under construction and will be reported in a future publication.

The high image bandwidth provided by MA-FPM offers a route to video-rate gigapixel microscopic imaging, which has not previously been possible. This is a primary requirement for imaging fast moving cell cultures [10]. MA-FPM retains all the advantages of conventional single-aperture FPM and can be very flexible in system design. For example, sample-specific selective spatial-frequency sampling was proposed to reduce the number of LEDs used, which can also be implemented to reduce the number of cameras used [51,52]. Importantly, MA-FPM is compatible with schemes involving multiplexed-illumination to further reduce the data acquisition time, indeed it provides the possibility for snapshot imaging, with sub-micron resolution over fields of view measured in square centimeters. MA-FPM can also be easily adapted for various FP based imaging modalities [45,53]. Camera-array based imaging is becoming established in macroscopic photography and surveillance to improve the image resolution (for aliased imaging systems) and to acquire wide-FoV images, but this is the first time a camera array has been reported in microscopy to improve the spatial and temporal resolution of microscopes.

**Funding**. This research is funded by The Scottish Universities Physics Alliance (SUPA).

**Acknowledgment**. We thank Dr. Paul Zammit for his help in using the camera and the LED array.

See Supplement 1 for supporting content.

# Supplement 1

## 1. Geometrical configuration of MA-FPM

The important parameters are shown in Fig S1. The lens-makers' formula determines the object and detector distances $u_c$ and $v_c$ and the magnification, $m$ of the system is chosen to satisfy the Nyquist sampling criteria. The displacement of the LED array from the object $s_d$ depends on the objective-lens NA, the LED separation $s_s$ and the overlap in frequency space. An optimal overlap of 60% is desired in all experimental configurations. For both the planar and Scheimpflug MA-FPM configurations, the period of the lens array pupil, $L_p$, and the number of LEDs determines the synthetic NA. They are related as following:

$$L_p = n \left( \frac{s_s u_c}{s_d} \right), \quad \text{(S1)}$$

where, $n$ is the LED position number from the centre of the array. For e.g., in a 5x5 LED array $n$ is 2. It is assumed that the LED array is odd numbered, and the central LED is on the optical axis of the central objective lens. This is chosen such that the LEDs in the array sample the spatial frequencies in the gaps between the objective lenses with the essential redundancy.

### A. Planar MA-FPM configuration

In the planar MA-FPM configuration, $m$ and $L_s$ are sufficient to determine positions of all the objective lenses and the detectors. $u_c$ and $v_c$ are calculated from the lens makers' formula, $L_s$ is equal to the period $L_p$ calculated using equation S1 and $D_s$ is calculated as $(1+m)L_s$.

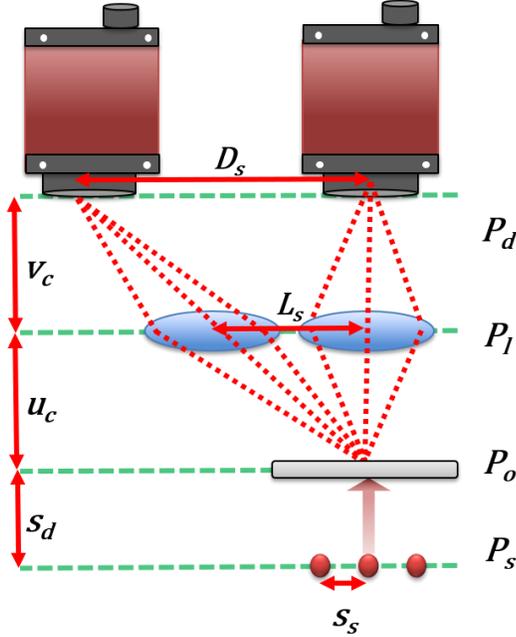

**Fig. S1** Planar MA-FPM configuration. $s_s$: Separation between the LEDs, $s_d$: LED array distance from the object, $u_c$: Lens array distance from the object, $v_c$: Detector array distance from the lens array, $L_s$: Separation between the objective lenses, $D_s$: Separation between the detectors, $P_s$: LED array plane, $P_o$: Object plane, $P_l$: Lens array plane, $P_d$: Detector array plane.

### B. Scheimpflug MA-FPM configuration

In this configuration, the lenses are tilted so the separation between the lenses is different compared to the planar configuration as observed in Fig. S2. Also, the objective lenses and the detectors have an additional variable, the angle of tilt to meet the Scheimpflug condition. The magnification can be varied between the central imaging system and the off-axis imaging systems. Hence, the magnification of the central imaging system is defined as $m_c$ and the magnification of the off-axis imaging system is defined as $m_o$. It is assumed that all the objective lenses have same focal length $f$. To determine positions of all the objective lenses and the detectors, the magnifications and the lens tilt angle are required. Magnifications are chosen to satisfy the imaging system Nyquist criteria and the lens tilt angle is calculated as follows:

$$\theta_l = \tan^{-1}\left( n \frac{s_s}{s_d} \right) \quad \text{(S2)}$$

This tilt angle is directly related to the lens array pupil period $L_p$. The central imaging system parameters $u_c$ and $v_c$ are calculated using the central imaging system magnification $m_c$ and the lens makers' formula. The remaining parameters to calculate the positions of the objective lenses and the detectors can be derived from trigonometrical arguments. They are given by the following equations:

$$L_s = \frac{1+m_o}{m_o} f \sin\theta_l, \quad \text{(S3)}$$

$$L_h = \frac{1+m_o}{m_o} f \cos\theta_l, \quad \text{(S4)}$$

$$D_s = \frac{(1+m_o)^2}{m_o} f \sin\theta_l, \quad \text{(S5)}$$

$$D_h = \frac{(1+m_o)^2}{m_o} f \cos\theta_l, \quad \text{(S6)}$$

$$\theta_d = \tan^{-1}\left( \frac{(1+m_o)\cos\theta_l \sin\theta_l}{1-(1+m_o)\sin^2\theta_l} \right). \quad \text{(S7)}$$

## 2. Field propagation using Fresnel diffraction integral

In this section, we discuss implementation of the Fresnel diffraction integral to propagate the optical field between various planes in the reconstruction procedure. The sampling requirement in different planes is also discussed.

### A. Diffraction integral with Fresnel approximation

The Fresnel diffraction integral that relates the field in the plane of interest $(x,y)$ from a plane of origin $(\xi,\eta)$ can be written in the form of a Fourier transform [35]:

$$U(x,y) = \frac{e^{jkz}}{j\lambda z} e^{j\frac{k}{2z}(x^2+y^2)} \int\int_{-\infty}^{\infty} U(\xi,\eta) e^{j\frac{k}{2z}(\xi^2+\eta^2)} e^{-j\frac{2\pi}{\lambda z}(x\xi+y\eta)} d\xi d\eta,$$

$$\quad \text{(S8)}$$

which can be rewritten in the form:

$$U(x,y) = e^{j\frac{k}{2z}(x^2+y^2)} \mathbb{F}\{U(\xi,\eta) e^{j\frac{k}{2z}(\xi^2+\eta^2)}\}, \quad \text{(S9)}$$

where the scalar factor is ignored for simplicity. Here multiplication of the complex field $U(\xi,\eta)$ with a quadratic phase factor accounts for near-field diffraction. This resultant field is then multiplied by another quadratic phase factor to give the propagated field in the plane $U(x,y)$.

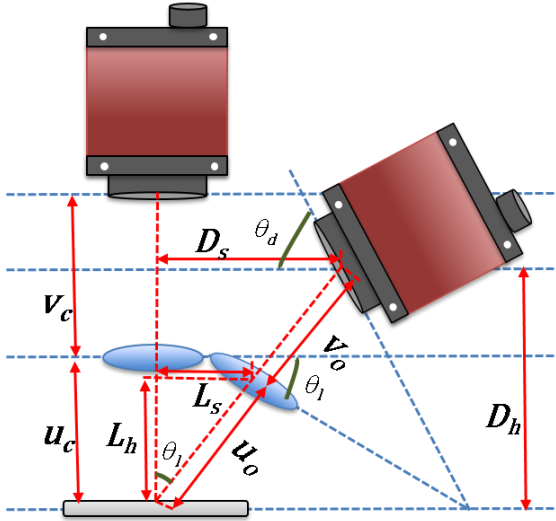

**Fig. S2** Scheimpflug MA-FPM configuration. $u_c$: Distance between the object and the central objective lens, $v_c$: Distance between the central objective lens and the central detector, $u_o$: Distance between the object and the off-axis objective lens, $v_o$: Distance between the off-axis objective lens and the off-axis detector, $\theta_l$ is the off-axis objective lens tilt angle, $L_h$ is the displacement of the off-axis objective lens from the object plane, $L_s$ is the separation between the objective lenses, $\theta_d$ is the off-axis detector tilt angle, $D_h$ is the displacement of the off-axis detector from the object plane, $D_s$ is the separation between the detectors.

Here the quadratic phase multiplied in both planes must be sampled according to the Nyquist frequency to avoid aliasing; i.e., we require that the sample period is such that [54]:

$$\delta < \frac{\lambda z}{2(\xi_{max} + \eta_{max})}, \quad (S10)$$

where $\delta$ is the pixel period, $\lambda$ is the wavelength, $z$ is the separation between the planes, $\xi_{max}$ and $\eta_{max}$ are the maximum extensions of the complex field in transverse directions.

**Propagation from the object plane to the lens plane**

Denoting the complex field in the object plane as $O(x,y)$, the complex field in the lens plane, $L(\xi,\eta)$ is given by:

$$L(\xi,\eta) = e^{j\frac{k}{2u}(\xi^2+\eta^2)}\mathbb{F}\{O(x,y)e^{j\frac{k}{2u}(x^2+y^2)}\} \quad (S11)$$

where $u$ is the propagation distance from the object plane to the lens plane, $k$ is the wavenumber, $(x,y)$ are coordinates in the object plane and $(\xi,\eta)$ are coordinates in the lens plane.

**Propagation from the lens plane to the detector plane**

The complex field in the detector plane $D(\alpha,\beta)$ is given by

$$D(\alpha,\beta) = e^{j\frac{k}{2v}(\alpha^2+\beta^2)}\mathbb{F}\{L(\xi,\eta)e^{j\frac{k}{2v}(\xi^2+\eta^2)}e^{j\frac{-k}{2f}(\xi^2+\eta^2)}\} \quad (S12)$$

where $v$ is the propagation distance from the lens plane to the detector plane and $f$ is the focal length of the lens. In this equation an additional quadratic phase factor is multiplied corresponding to the lens transfer function [21].

To propagate the field back from the detector plane to the lens plane and thence to the object plane similar equations to S12 and S11 can be used respectively, but with a complex-conjugate quadratic phase distribution. This would cancel out the requirement of multiplying the quadratic phase factor in the detector plane – relaxing the sampling requirement – since we are interested only in the amplitude distribution.

It should also be noted that in the lens plane three quadratic phase factors are multiplied before applying the Fast Fourier transform: (1) from propagating the field from the object plane to the lens plane from equation S11, (2) from the quadratic phase multiplied before propagating the field from the lens plane to the detector plane, and (3) from the lens transfer function as shown in equation S12 [35]. This quadratic phase is for an ideal lens without aberrations, but additional aberrations can also be incorporated. Multiplying these three quadratic phase factors will tend to cancel leaving the phase of only the aberrations. Hence the spatial sampling frequency is not a concern in the lens plane.

Nyquist sampling in the object plane requires careful consideration, however, from equation S10 it can be calculated that the sample interval in the object plane should be less than 10 microns, which is easily satisfied in our MA-FPM reconstruction since the pixel size in the object plane is around 0.2 microns.

## 3. Calibration results

For calibration of the planar MA-FPM geometry and phase aberrations of the objective lenses, an FPM dataset is recorded using a set of circular LEDs with a diameter of 19 LEDs that are unique to each objective lens. The LED sets are chosen such that the similar set of spatial-frequency bands in the reconstructed high-NA images are recorded by each camera. These datasets are processed using the MA-FPM reconstruction procedure described in the main text, except the number of cameras in this case would be one. Nine such datasets, each corresponding to one of the lenses in the array, are processed to obtain the results such as are shown in Fig.S5. For each camera, we make an initial estimate of the phase aberrations present in the system using a *Zemax* model of the optics; these are shown in column four of Fig.S5. It can be observed in the results that the aberrations vary with the lens position, in the lens positions 2, 4, 6, 8, coma and astigmatism have symmetry about vertical or horizontal axes whereas for lenses 1, 3, 7, 9 the symmetry axes are at ±45° and have greater magnitudes as would be expected from the relative off-axis locations of the lenses. The increase in magnitude of phase aberrations with off-axis imaging manifests in the relative aberrations of the low-NA images in column 1: ranging from the least aberrated camera-1 image to the most highly aberrated images for cameras 1, 3, 7 and 9. Following reconstruction using the pupil recovery of the algorithm a new updated phase estimate is generated as shown in the fifth column of Fig.S5. This pupil phase estimate is a better approximation of the true aberrations present in the actual system to be used in the MA-FPM reconstruction. It can be seen in Fig.S3 that if the *Zemax* estimated pupils are used instead of the recovered pupils shown in fifth column of Fig. S5, the recovered image suffers from artefacts. The group seven element four recovery has artefacts when only *Zemax* estimated pupils are used, which is not the case when pupils recovered from the calibration procedure are used.

The reconstructed high-NA images (column 2) in Fig. S5 look similar since the aberrations are corrected during the reconstruction. These images now differ only in translation because of the errors in relative positions of the detectors and can be registered easily by calculating the appropriate translation between the images. We take the central camera (5) image as a reference to which the other eight images are registered. A translation of eight pixels in the high-NA image corresponds to one-pixel translation in the low-NA image because of the resizing factor. These calculated translations are applied to the low-NA images to correct for the offsets of the detectors. Once these translations are applied, the high-NA images will be registered as seen in the third column of the figure. These translations are used to register the MA-FPM datasets.

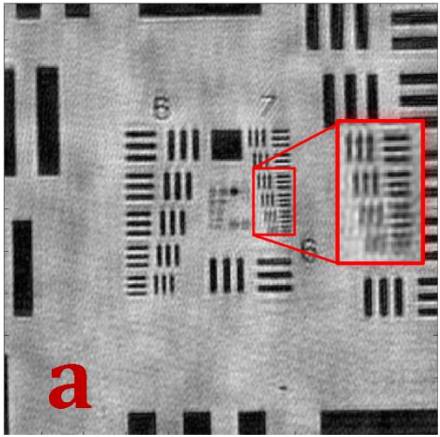

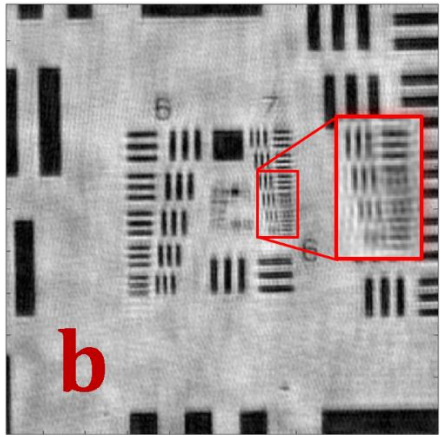

**Fig.S3** Comparison between MA-FPM reconstruction with (a) and without (b) calibrated pupil

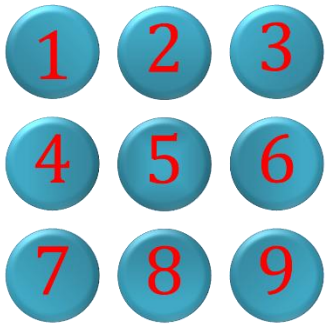

**Fig. S4.** Lens positions numbering for reference to the next figure

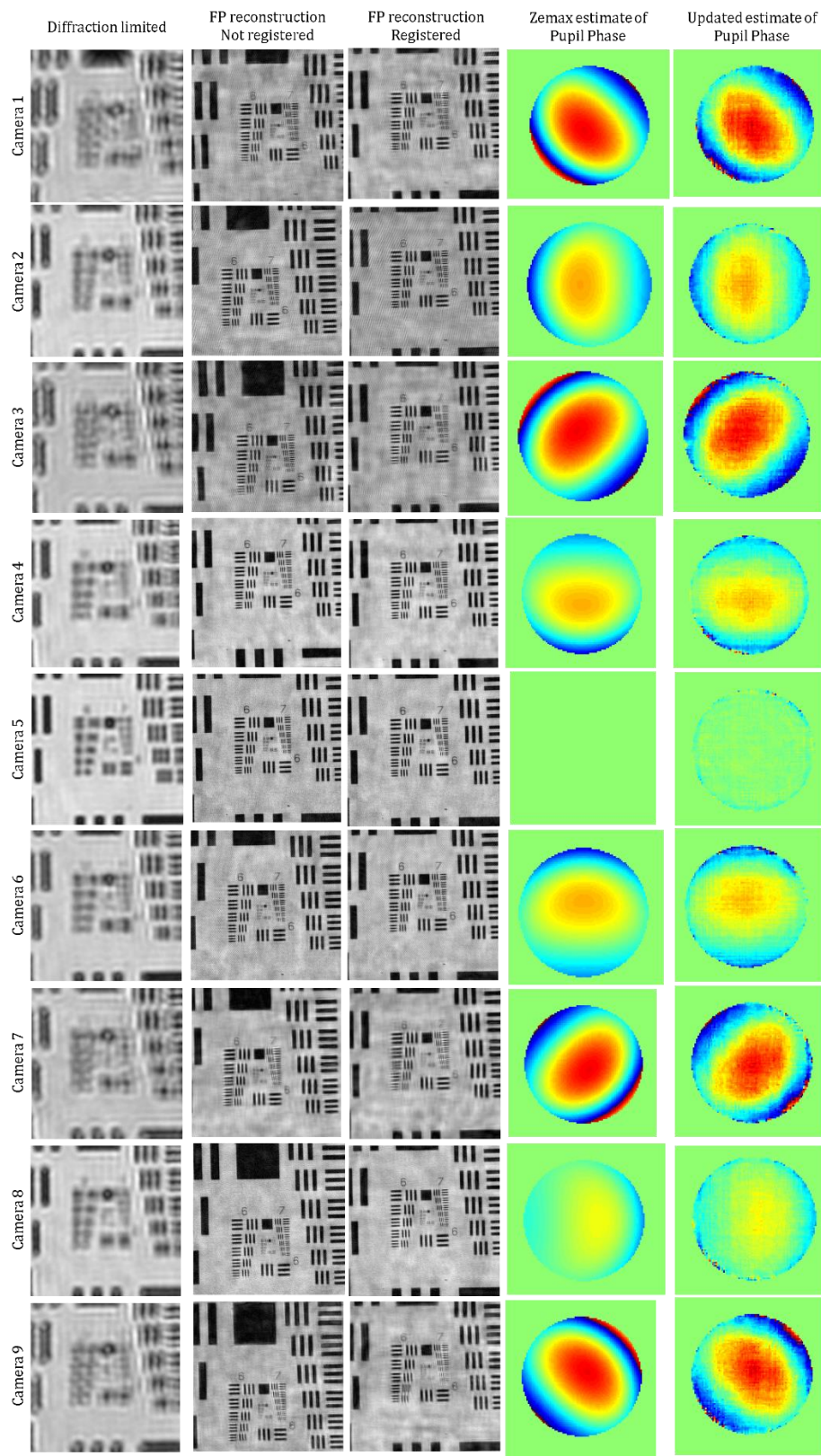

**Fig. S5** FPM reconstruction of calibration datasets. For pupil phase images the phase range is from −pi (blue) to +pi (red).